\documentclass[twocolumn,showpacs,preprintnumbers,%
amsmath,amssymb,prd]{revtex4}

\usepackage{dcolumn}
\usepackage{bm}
\usepackage{longtable}

\input amssym.def
\input amssym
\newcommand{\be}{\begin{equation}}
\newcommand{\ee}{\end{equation}}
\begin{document}

\title{A $T_0$-discrete universe model
with five low-energy fundamental
interactions\\ and the coupling constants hierarchy}

\author{Vladimir N. Efremov}%
 \email{efremov@udgserv.cencar.udg.mx}
\affiliation{Mathematics Department, CUCEI,
University of Guadalajara, Guadalajara, Jalisco, Mexico.}

\author{Nikolai V. Mitskievich}%
 \email{nmitskie@udgserv.cencar.udg.mx}
\affiliation{Physics Department, CUCEI, University
of Guadalajara, Guadalajara, Jalisco, Mexico.}

\date{\today}

\begin{abstract}
A quantum model of universe is constructed in which values of
dimensionless coupling constants of the fundamental interactions
(including the cosmological constant) are determined via certain
topological invariants of manifolds forming finite ensembles of 3D
Seifert fibrations. The characteristic values of the coupling
constants are explicitly calculated as the set of rational numbers
(up to the factor $2\pi$) on the basis of a hypothesis that these
values are proportional to the mean relative fluctuations of
discrete volumes of manifolds in these ensembles. The discrete
volumes are calculated using the standard Alexandroff procedure of
constructing $T_0$-discrete spaces realized as nerves
corresponding to characteristic canonical triangulations which are
compatible with the Milnor representation of Seifert fibered
homology spheres being the building material of all used 3D
manifolds. Moreover, the determination of all involved homology
spheres is based on the first nine prime numbers ($p_1=2, \dots,
p_9=23$). The obtained hierarchy of coupling constants at the
present evolution stage of universe well reproduces the actual
hierarchy of the experimentally observed dimensionless low-energy
coupling constants.
\end{abstract}

\pacs{04.20.Gz,98.80.Bp,98.80.Hw}

\maketitle

Almost half a century ago Wheeler \cite{1} has put forward the
idea of the quantum spacetime topology fluctuations. This idea had
been further developed by Hawking \cite{2} using the path integral
techniques in the Euclidean quantum gravity. Topology fluctuations
are still a crucial element of the wormhole (or, baby-universe)
mechanism for vanishing of the cosmological constant \cite{3}.
Some efforts were also made to show that the entire effect of the
spacetime topology fluctuations were to modify all fundamental
coupling constants in physics and to provide a probability
distribution for them (see, {\em e.g.}, \cite{4}). We develop here
similar ideas, but now in terms of the $T_0$-discrete spacetime
realized as the inverse spectrum of nerves of the sequence of
triangulations \cite{5} ({\em a discrete canonical approach}). This
makes it possible to quantitatively deduce the hierarchy of {\em
dimensionless low-energy coupling} (DLEC) constants of the known
fundamental interactions, as well as the evolution of these
``constants'' while the universe goes through a sequence of
``inflationary'' stages. To this end we use a finite ensemble of
topological spaces ${\mathcal M}$ which describes a spectrum of
the topology fluctuations of three-dimensional spatial sections.
Just the numerical characteristics of these fluctuations determine
magnitudes of the coupling constants. The principal constructive
elements in building the ensemble ${\mathcal M}$ are the {\em Seifert
fibered homology spheres} (Sfh-spheres) characterized by
collections of three mutually prime numbers. Moreover, the
determination of all involved Sfh-spheres is based on the first
nine prime numbers ($p_1=2, \dots, p_9=23$). So one can trace a
relation between the coupling constants hierarchy of the
fundamental interactions and the sequence of the prime numbers in
the beginning of the set of positive integers $\Bbb N$.

\begin{center}
{\bf {\boldmath $T_0$}-discrete spaces}
\end{center}

By an Alexandroff space \cite{6} we mean a topological space every point
of which has a minimal neighborhood (assignment of minimal neighborhoods
of all points fixes the topology of every Alexandroff's space). We shall consider
here only the Alexandroff spaces with the $T_0$ separability axiom which we
call $T_0$-discrete spaces. Nerves of triangulations of smooth compact
manifolds represent an important example of the $T_0$-discrete spaces \cite{6}.
Recall that any smooth compact three-dimensional manifold $X$ admits a {\em finite
closed partition} $\tau=\{T_1,\dots,T_N\}$ consisting of closed 3D-simplices
(tetrahedra) $T_i$ such that $\text{dim}(T_i\cap T_j)<3$, for $\forall ~ i,j=1,\dots,
N, ~ i\ne j$. This partition is called the ({\em canonical}) c-triangulation. We call
the {\em nerve} of a c-triangulation $\tau$ the topological space (simplicial complex)
$X_\tau$ whose points (simplices) are collections $\{T_{i_0},\dots,T_{i_q}\}=:
x_{i_0\dots i_q}$ of elements of $\tau$ for which $T_{i_0}\cap\dots\cap T_{i_q}
\ne\emptyset$, while the minimal neighborhood of any point $x_{i_0\dots i_q}$
is defined as the set of points $x_{j_0\dots j_p}\in X_\tau$ such that
$T_{j_0}\cap\dots\cap T_{j_p}\subseteq T_{i_0}\cap\dots\cap T_{i_q}$.
Any nerve is a $T_0$-discrete space \cite{6}. In the topology defined by
the last relation, not all points of the nerve are closed since $X_\tau$ is {\em not} a
Hausdorff space. The ({\em closed}) c-points of $X_\tau$ are those and only those which
correspond to tetrahedra, {\em i.e.} $x_{i_0}=\{T_{i_0}\}$. Later on we shall call
{\em discrete volume} of a manifold $X$, corresponding to c-triangulation, the number
$N$ of tetrahedra in the triangulation $\tau$, or equivalently, the number of
c-points in the nerve $X_\tau$.

\begin{center}
{\bf Seifert fibered homology spheres}
\end{center}

The basic structural elements of which we construct our cosmological model,
are Sfh-spheres. We shall make use of
explicit analytic description of Sfh-spheres with three exceptional fibers
\cite{7}. Let $\Sigma (a_1,a_2,a_3)=:\Sigma (\underline{a})$ be
the smooth compact three manifold obtained by intersecting the complex
algebraic Brieskorn surface ${z_1}^{a_1}+{z_2}^{a_2}+{z_3}^{a_3}=0$
($z_i \in \Bbb C _i$) with the unit five dimensional sphere
$|z_1|^2+|z_2|^2+|z_3|^2=1$, where $a_1, a_2, a_3$ are pairwise coprime integers,
$a_i > 1$.  There exists a unique Seifert fibration of this manifold with
unnormalized Seifert invariants \cite{8}:
$(a_i,b_i)$ subject to $e(\Sigma (\underline{a}))=
\sum_{i=1}^{3} b_i/a_i =1/a$, where $a=a_1 a_2 a_3$ and $e(\Sigma
(\underline{a}))$ is the well known topological invariant of a Sfh-sphere
also called its Euler number.

To construct our model of universe we need a specific family of
Sfh-spheres which would be defined in following way. First, the derivative
of a Sfh-sphere $\Sigma (\underline{a}):=\Sigma (a_1,a_2,a_3)$ can be
defined as a Sfh-sphere
\be \label{1}
\Sigma^{(1)}(\underline{a}):=\Sigma (a_1,a_2a_3,a+1)\equiv
\Sigma (a^{(1)}_1,a^{(1)}_2,a^{(1)}_3).
\ee
The Euler number of this Sfh-sphere is
$e(\Sigma^{(1)}(\underline{a}))=1/a^{(1)}$
where $a^{(1)}=a^{(1)}_1a^{(1)}_2a^{(1)}_3=a(a+1)$.
By induction, we define the derivative $\Sigma^{(l)}(\underline{a})=
\Sigma(a^{(l)}_1,a^{(l)}_2,a^{(l)}_3)$ of
$\Sigma(\underline{a})$ of any
order $l$. In particular, there holds the recurrent relation
\be \label{2}
a^{(l)}=a^{(l-1)}\left(a^{(l-1)}+1\right)
\ee
for a product of three Seifert invariants $a^{(l)}
=a^{(l)}_1a^{(l)}_2a^{(l)}_3$.
Second, we define a sequence of Sfh-spheres which we shall
call {\em primary sequence}. Let $p_i$ be a prime number
being the $i$th in the set of the positive integers $\Bbb N$,
{\em e.g.}, $p_1=2, ~ p_2=3,\dots$.
The primary sequence of Sfh-spheres is defined as
\be \label{3} \{\Sigma(q_i,p_{i+1},p_{i+2})|i\in{\Bbb N}\}
\ee where $q_i=p_1\cdots p_i$.
Finally, to the end of constructing our model of universe,
we include in this sequence as its first two terms the
usual {\em three-dimensional spheres} $S^3$ {\em with Seifert's
fibrations} (Sf-spheres) determined by the mappings
$h_{pq}:S^3\rightarrow S^2$ \cite{9}.
Recall that $S^3=\left\{(z_1,z_2)
\left| |z_1|^2+|z_2|^2=1\right.\right\}$ and $z^p_1/z^q_2
\in{\Bbb C}\cup\{\infty\}\cong S^2$.
We denote these two Sf-spheres
as $\Sigma(1,1,2)$, $p=1, q=2$ and $\Sigma(1,2,3)$,
$p=2, q=3$. In these notations we use an additional third
number (unit) which corresponds to an arbitrary regular
fiber. This will enable us to take derivatives of Seifert
fibrations on $\Sigma(1,1,2)$ and $\Sigma(1,2,3)$ by the
same rule (\ref{1}) as for other members of the
sequence (\ref{3}).

Now we form the family of manifolds corresponding to the
first primary Sfh-spheres and their derivatives up to the
fourth order,
\be \label{4}
\{\Sigma^{(l)}(q_{i-1},p_i,p_{i+1})|i\in\overline{0,8},
l\in\overline{0,4}\},
\ee
where $\overline{0,n}$ is the integer numbers interval from
0 to $n$. Note that the subfamily corresponding to $i=0,1$
is built of the ordinary spheres $S^3$ with fixed Seifert
fibrations. In order to include the
Sf-spheres $\Sigma(1,1,2)$ and $\Sigma(1,2,3)$ in this
family, one has to put $q_{-1}=q_0=p_0=1$. {\em E.g.},
for the well known Poincar\'e homology sphere $\Sigma(p_1,
p_2,p_3)=\Sigma(2,3,5)$, the sequence of derivatives is
$$ \begin{array}{l}
\Sigma^{(1)}(2, 3, 5) = \Sigma(2, 15, 31),\\
\Sigma^{(2)}(2, 3, 5) =\Sigma(2, 465, 931),\\
\Sigma^{(3)}(2, 3, 5) = \Sigma(2, 432915, 865831),\\
\Sigma^{(4)}(2, 3, 5) = \Sigma(2, 374831227365, 749662454731).
\end{array}
$$

Having done calculations of the Euler numbers of Seifert
structures of Sf- and Sfh-spheres in the family (\ref{4}),
we find that for the subfamily
\be \label{5}
\left\{\left. \Sigma^{(l)}=\Sigma^{(l)}(q_{2l-1},p_{2l},p_{2l+1})
\right|l\in\overline{0,4}\right\} \ee the Euler numbers (multiplied
by $2\pi$) reproduce fairly well the real hierarchy of
DLEC constants of fundamental interactions, see Table \ref{tab:1}.
\begin{table}[h]
\caption{ \label{tab:1} Euler numbers {\it vs.} experimental DLEC constants.}
\begin{ruledtabular}
\begin{tabular}{llll}
$l$ & {$e\left(  \Sigma^{l}\right) $} & {Interaction} &
{$\alpha_\text{exper}$\footnote{Reference \cite{11}.}}\\ \hline
$0$ & $3.14$ & strong & $1$\\ \hline
$1$ & $6.78\times10^{-3}$ & electromagnetic &
$7.20\times10^{-3}$\\ \hline
$2$ & $2.20\times10^{-13}$ & weak &
$3.04\times10^{-12}$\\ \hline
$3$ & $1.36\times10^{-45}$ & gravitational\footnote{Normalized
with respect to the electron mass.} &
$2.73\times10^{-46}$\\ \hline
$4$ & $1.67\times10^{-133}$ & cosmological & $<10^{-120}$
\end{tabular}
\end{ruledtabular}
\end{table}

The agreement of the calculated hierarchy of the DLEC constants and
the experimental data suggests the idea to construct a model of
universe glued up of Sf- and Sfh-spheres using the connected sum
operation (compact locally homogeneous universes with spatial
sections homeomorphic to Seifert fibrations were considered in
\cite{10}). To this end we primarily have to reduce and
reparametrize the family (\ref{4}). First, in accordance with
(\ref{5}), we eliminate the Sf- and Sfh-spheres with odd numbers
$i$ introducing a new parameter $n\in \overline{0,4}$ related to
$i$ as $i=2n$. Then (in certain cases) it  is convenient also to
use another parameter $t=n-l, ~ t\in \overline{-4,4}$. The
resulting family of Sf- and Sfh-spheres is \be   \label{6}
\left\{\left.\Sigma^{(l)}_n:=\Sigma^{(n-t)}(q_{2n-1},p_{2n},
p_{2n+1})\right|n\in\overline{0,4},t\in\overline{-4,4} \right\},
\ee which contains (\ref{5}) as a subset for $t=0$, {\em i.e.}
when $n=l$. Parameter $t$ in our model is the discrete
cosmological ``time'', $t=0$ labelling the present state of the
universe where an observer can determine the DLEC constants
$\alpha^{(n)}_n$ of the five ($n\in\overline{0,4}$) fundamental
interactions (see Table \ref{tab:1}). The relation (\ref{2})
readily yields good estimates of the DLEC constants
$\alpha^{(n)}_n= 2\pi e\left(\Sigma^{(n)}_n\right)\simeq
2\pi(q_{2n+1})^{-2^n}$. Remember that $q_{2n+1}=p_1\cdots
p_{2n+1}$ is product of the first $2n+1$ prime numbers in ${\Bbb
N}$. Note that for $n=5$ there would be $\alpha^{(5)}_5\simeq
1.4\cdot 10^{-357}$ which is too small to be identified with a
certain experimentally determined coupling constant, thus we put
$n,l\in\overline{0,4}$.

Though this approach leads to a hierarchy of the DLEC constants,
it yields neither a description of the evolution of universe, nor
other its features, therefore we pass to framing a more
constructive universe model glued of Sf- and Sfh-spheres in
accordance with the method described in \cite{5} a particular
case of which we present below. To this end we first have to
calculate the discrete volumes of Sf- and Sfh-spheres from the
family (\ref{6}).

\begin{center}
{\bf Canonical characteristic triangulations
}
\end{center}

We now have to find such minimal c-triangulations of a Sfh-sphere
$\Sigma(a_1,a_2,a_3)$ which are compatible with its Milnor
representation as
a $a_3$-fold cyclic branched covering of $S^3$ branched along a
torus knot $K(a_1,a_2)$ \cite{7}. We call this triangulation the
{\em characteristic canonical triangulation} (chc-triangulation or $\tau_c$).
The formula giving the number of tetrahedra in $\tau_c$, or the
number of c-points in the corresponding nerve, is invariant
under permutations of Seifert invariants $a_1,a_2,a_3$ since the
Milnor representation is symmetric under their permutations.

A c-triangulation is said to be compatible with the Milnor
representation of $\Sigma(a_1,a_2,a_3)$ if the knot $K(a_1,a_2)$
consists of one-dimensional simplices of the triangulation induced by
$\tau_c$. To find the chc-triangulation, consider in $S^3$ a torus
$T^2$ with a torus knot $K(a_1,a_2)$ on it. The simplest triangulation
of this torus when the knot $K(a_1,a_2)$ is a 1-cycle, contains $2a_1a_2$
triangles. These triangles are now used as 2-faces in c-triangulation
of the sphere $S^3$ divided by $T^2$ into two solid tori $ST_1$ and
$ST_2$. This triangulation of the torus $T^2$ induces
the minimal triangulation of each of the two solid tori, which consists of
$8a_1a_2$ tetrahedra. Thus the total number of tetrahedra in the
c-triangulation of $S^3$ is $16a_1a_2$. Lifting this triangulation to
$\Sigma(a_1,a_2,a_3)$, we get $16a_1a_2a_3=16a$ tetrahedra in the
chc-triangulation.

Note that the Milnor covering $\pi:\Sigma(a_1,a_2,a_3)\to S^3$ is a
simplicial one \cite{6} in the sense that chc-triangulation of
$\Sigma(a_1,a_2,a_3)$ with the number of tetrahedra $16a$ is
 a covering one of either of c-triangulations of $S^3$ with the numbers
of tetrahedra $16a_1a_2$, or $16a_1a_3$, or $16a_2a_3$; we shall
call them chc-triangulations of $S^3$. Therefore when at the $n$th
level the gluing is performed of either $\Sigma(a_1,
a_2,a_3)$ or $S^3$, $\Sigma(a_1,a_2,a_3)$ is taken in
chc-triangulation with $N(\Sigma)=16a$ tetrahedra, while $S^3$, in
any one of its three chc-triangulations, so that the mean number of
tetrahedra now is $N(S)=(16/3)(a_1a_2+a_1a_3+a_2a_3)$.
Here $N(\Sigma)$ and $N(S)$ are measures of discrete volume of
the manifolds $\Sigma(a_1,a_2,a_3)$ and $S^3$ respectively.
Note that at the $n$th level $\Sigma(a_1,a_2,a_3)=\Sigma^{(l)}(q_{2n-1},
p_{2n},p_{2n+1})$. Numbers of tetrahedra in the $n$th-level
triangulations are written as $N^{(l)}_n(\Sigma)$ and $N^{(l)}_n(S)$
(the gluing is performed under the condition $l=$ const).

\begin{center}
{\bf Universe with one fundamental interaction}
\end{center}

In our model every fundamental interaction will be characterized
by a pair of discrete parameters $(n,l)$,  where
$n,l\in\overline{ 0,4}$. Any $(n,l)$-inter\-action
is related to an ensemble ${\mathcal M}^{(l)}_n$ of topological spaces
$M^{(l)}_n(R)$ which are interpreted as a set of admissible
spatial sections of a ``universe'' involving only one fundamental
interaction. Such a ``universe'' will be called
$(n,l)$-universe.

Let the ensemble ${\mathcal M}^{(l)}_0$ consist of merely one manifold
$M^{(l)}_0=\Sigma^{(l)}_0$, a Sf-sphere from the family (\ref{6}).
The chc-triangulation of this Sf-sphere contains
$N^{(l)}_0(\Sigma)$ tetrahedra, thus the discrete volume of
$M^{(l)}_0$ is equal to $V^{(l)}_0=N^{(l)}_0(\Sigma)$. The
ensemble ${\mathcal M}^{(l)}_1:=\left\{
\left.M^{(l)}_1(R)\right|R\in\overline{0,V^{(l)}_0}\right\}$
consists of components $M^{(l)}_1(R)$ every one of which is
obtained from the manifold $M^{(l)}_0$ by $R$-fold application of
the connected sum operation involving the Sfh-sphere
$\Sigma^{(l)}_1$ and $\left(V^{(l)}_0-R\right)$-fold application
of the same operation involving $S^3$. Strictly speaking, one has
to remove $R$ tetrahedra from the manifold $M^{(l)}_0$ and to
attach instead of them $R$ Sfh-spheres $\Sigma^{(l)}_1$ (from each
of the latters one tetrahedron is also supposed to be removed). A
similar procedure has to be performed using
$\left(V^{(l)}_0-R\right)$ chc-triangulated spheres. The obtained
manifold $M^{(l)}_1(R)$ has discrete volume (tetrahedra number in
the resulting triangulation) \be \label{7}
N^{(l)}_1(R)=RN'^{(l)}_1(\Sigma)+\left(V^{(l)}_0-R\right)N'^{(l)}_1(S) \ee
where $N'^{(l)}_1(\Sigma)=N^{(l)}_1(\Sigma)-1$ and
$N'^{(l)}_1(S)=N^{(l)}_1(S)-1$; ($-1$) appeared here due to the
removal of a tetrahedron from both $\Sigma^{(l)}_1$ and $S^3$ when
the connected sum was applied.
\begin{table*}
\caption{ \label{tab:2} Mean discrete volumes of $(n,l)$-universes.}
\begin{tabular}[c]{|c|c|c|c|c|c|c|c|c|c|}\hline
$_{n}\diagdown^{t}$ & 4 & 3 & 2 & 1 &
0 & $-1$ & $-2$ & $-3$ & $-4$\\\hline
0 &  &  &  &  & $3.2\times10^1$ & $9.6\times10^1$ &
$6.7\times10^2$ & $2.9\times10^4$ & $5.2\times10^7$\\\hline
1 &  &  &  &$1.0\times10^4$ & $8.6\times10^5$ &
$ 5.4\times10^9$ & $2.0\times10^{17}$ & $2.7\times10^{32}$ & \\\hline
2 &  &  & $2.1\times10^8$ &$3.7\times10^{13}$ &
$ 1.3\times10^{24}$ &
$1.3\times10^{45}$ & $1.5\times10^{87}$ &  & \\\hline
3 &  &$8.9\times10^{14}$ &$7.8\times10^{25}$ & $6.8\times10^{47}$ &
$4.9\times10^{91}$ &$2.5\times10^{179}$
&  &  & \\\hline
4 & $1.6\times10^{24}$ & $3.1\times10^{42}$ &
$ 1.3\times10^{82}$ &
$2.4\times10^{159}$ & $7.5\times10^{313}%
$ &  &  &  & \\ \hline
\end{tabular}
\end{table*}
Note that the discrete volume $N^{(l)}_1(R)$ does not depend on
the choice of concrete $R$ tetrahedra instead of which the
Sfh-spheres are pasted. Let us suppose that the probabilities of
pasting an Sfh-sphere and the usual $S^3$, are the same
($p_\Sigma=p_S=1/2$), and characterize the ensemble ${\mathcal
M}^{(l)}_1$ by the {\it mean discrete volume} (md-volume) of the
$(1,l)$-universe $V^{(l)}_1=[
\sum^{V^{(l)}_0}_{R=0}w^{(l)}_1(R)N^{(l)}_1(R) ]$ where $[\dots]$
is the integer part of of the number in these brackets, while
$w^{(l)}_1(R)=\binom{V^{(l)}_0}{R}/2^{V^{(l)}_0}$ is a particular
case of the well known binomial (Bernoulli) distribution which we
relate to the $(1,l)$-interaction.

By induction, the md-volume $V^{(l)}_n$ of the $(n,l)$-universe
can be found from $V^{(l)}_{n-1}$ (md-volume of the ensemble
${\mathcal M}^{(l)}_{n-1}:=\left\{\left.M^{(l)}_{n-1}(R)\right|
R\in\overline{0,V^{(l)}_{n-2}}\right\}$). It reads as
$$V^{(l)}_n=\left[ \sum^{V^{(l)}_{n-1}}_{R=0}w^{(l)}_n(R)N^{(l)}_n(R)
\right]$$ where
$w^{(l)}_n(R)=\binom{V^{(l)}_{n-1}}{R}/2^{V^{(l)}_{n-1}}$ is the
binomial distribution of the $(n,l)$-interaction which for large
$V^{(l)}_{n-1}$ tends to the Gaussian distribution. From the
properties of this distribution it follows that \be \label{8}
V^{(l)}_n=V^{(l)}_{n-1}\left[\left(N'^{(l)}_n(\Sigma)+N'^{(l)}_n(S)
\right)/2\right]. \ee The numerical results for the md-volumes of
all $(n,l)$-universes are given in Table \ref{tab:2} (up to the
second significant digit). Then one can calculate the Shannon
entropy $S^{(l)}_n=-\sum^{V^{(l)}_{n-1}}_{R=0} w^{(l)}_n(R)\ln
w^{(l)}_n(R)$ of the $(n,l)$-universe (and in a certain sense of
the fundamental interaction of $(n,l)$-type). The corresponding
calculations and discussion of results concerning the cosmological
evolution, will be published elsewhere.

We came to the hypothesis that even a ``universe'' with merely one
interaction should be characterized by an ensemble of topological
spaces (topologies) ${\mathcal
M}^{(l)}_n:=\left\{\left.M^{(l)}_n(R)\right|
R\in\overline{0,V^{(l)}_{n-1}}\right\}$. This mixed-state
representation makes the quantum-mechanical description more
adequate; moreover, in a universe with several fundamental
interactions the quantum-theoretical approach becomes imperative.
Let us associate the spacelike section topology $M^{(l)}_n(R)$
with the state vector $|n,l,R\rangle$ characterized by ``quantum
numbers'' $n,l\in\overline{0,4}, ~
R\in\overline{0,V^{(l)}_{n-1}}$. Then to the ensemble ${\mathcal
M}^{(l)}_n$ corresponds a collection of pure states (basis)
$B^{(l)}_n(R)=\left\{|n,l,R\rangle\left|R\in\overline{0,V^{(l)}_{n
-1}}\right.\right\}$ generating the Hilbert space ${\mathcal
H}^{(l)}_n$. Transition between state vectors is realized by
operators of creation and annihilation of Sfh-spheres
$\Sigma^{(l)}_n$, which we define using the ordinary (boson)
commutation relations and the following set of rules: \be
\label{9} \begin{array}{l}
a^{(l)+}_n|n,l,R\rangle= \sqrt{R+1}|n,l,R+1\rangle, ~
R<V^{(l)}_{n-1};\\
a^{(l)-}_n|n,l,R\rangle= \sqrt{R}|n,l,R-1\rangle, ~ R>0;\\
a^{(l)+}_n|n,l,V^{(l)}_{n-1}\rangle= |n+1,l,0\rangle;  \\
a^{(l)-}_n|n,l,0\rangle= |n-1,l,V^{(l)}_{n-2}\rangle;  \\
a^{(l)\pm}_{n'}|n,l,R\rangle=0=a^{(l')\pm}_{n}|n,l,R\rangle, ~
n\neq n', ~ l\neq l'.
\end{array} \ee
\underline{Observation 1}. Here, the first two relations show that
the operators $a^{(l)\pm}_{n}$ describe minimal (in our model)
topology changes of spatial sections of $(n,l)$-universe related
to joining ($a^{(l)+}_{n}$) and detaching ($a^{(l)-}_{n}$) a
wormhole with spatial section homeomorphic to the Sfh-sphere
$\Sigma^{(l)}_{n}$. Therefore these operators can be as well
interpreted as discrete time shift generators. Topology changes
involving Sfh-spheres are, mathematically speaking,
Siebenmann-type cobordisms \cite{12}, and they were considered
from the physical viewpoint in \cite{13}.\\
\underline{Observation 2}. The state vector $|n,l,0\rangle$
describes the vacuum state of the $(n,l)$-universe. It is
associated with the spatial section $M^{(l)}_{n}(0)$ which does
not contain Sfh-spheres $\Sigma^{(l)}_{n}$, {\em i.e.} all
$V^{(l)}_{n-1}$ tetrahedra of the ($n-1$)st level
chc-triangulation are substituted by ordinary spheres $S^3$, but
with chc-triangulation of the $n$th level. Therefore by the action
of annihilation operator of $\Sigma^{(l)}_{n}$ on this state (the
fourth relation in (\ref{9})) occurs a transition to ``the nearest
lower'' state $|n-1,l,V^{(l)}_{n-2}\rangle$ now corresponding to
another, $(n-1,l)$, universe.\\
\underline{Observation 3}. Concerning the third relation in
(\ref{9}), note that any three-dimensional manifold $M$ is
homeomorphic to its connected sum with $S^3$, {\em i.e.} $M\cong
M\# S^3$. Hence the transition from $|n,l,V^{(l)}_{n-1}\rangle$ to
$|n+1,l,0\rangle$ due to the operator $a^{(l)+}_n$, does not
represent a topology change, but leads to a sharp growth of the
tetrahedra number in the manifold $M^{(l)}_n(V^{(l)}_{n-1})$
chc-triangulation. In fact this is a transition to the manifold
$M^{(l)}_{n+1}(0)$ homeomorphic to $M^{(l)}_n(V^{(l)}_{n-1})$, but
having considerably more tetrahedra in its chc-triangulation. We
relate this phenomenon to the cosmological inflation since in
terms of the $T_0$-discrete spaces (nerves of chc-triangulations)
this leads to a rapid growth of the c-points number. For example,
for ``cosmological'' interactions, {\em i.e.}
$(n,4)$-interactions, there are four inflationary stages. This
sequence of stages is described by a diagram of transitions
between the ensembles ${\mathcal
M}^{(4)}_0\stackrel{10^{24}}{\longrightarrow} {\mathcal
M}^{(4)}_1\stackrel{10^{53}}{\longrightarrow} {\mathcal
M}^{(4)}_2\stackrel{10^{89}}{\longrightarrow} {\mathcal
M}^{(4)}_3\stackrel{10^{128}}{\longrightarrow} {\mathcal
M}^{(4)}_4$ where numbers over the arrows give orders of magnitude
of the growth of the tetrahedra number by the transition from the
manifold $M^{(l)}_n(V^{(l)}_{n-1})$ of the ensemble ${\mathcal
M}^{(4)}_n$ to the manifold $M^{(l)}_{n+1}(0)$ of ${\mathcal
M}^{(4)}_{n+1}$. We associate the ``ordinary'' inflation with the
last transition characterized by the $10^{128}$-fold discrete
volume growth.
\begin{table*}
\caption{\label{tab:3}Dimensionless constants of fundamental
interactions}%
\begin{tabular}{|c|c|c|c|c|c|c|c|c|c|}\hline
$r^{(n-t)}_n$: & 1.4$\times10^{-35}$ & $9.4\times10^{-31}$ &
 $4.3\times10^{-21}$ & $8.8\times10^{-2}$ &
$3.7\times10^{37}$ & $8.9\times10^3$ & $7.8\times10^{-20}$
 & $1.2\times10^{-33}$ & $1.1\times10^{-39}$\\\hline
  $_{n}\diagdown^{t}$ & 4 & 3 & 2 & 1 &
  0 & $-1$ & $-2$ & $-3$ & $-4$\\\hline
0 &  &  &  &  & $1.1\times10^0$ & $6.4\times10^{-1}$
 & $2.4\times10^{-1}$ &
$3.7\times10^{-2}$ & $8.7\times10^{-4}$\\\hline
1 &  &  &  & $6.2\times10^{-2}$ & $6.8\times10^{-3}$ &
$8.5\times10^{-5}$ &
$1.4\times10^{-8}$ & $3.8\times10^{-16}$ & \\\hline
2 &  &  & $4.6\times10^{-4}$ & $1.0\times10^{-6}$ & $5.6
\times10^{-12}$ &
$1.7\times10^{-22}$ & $1.6\times10^{-43}$ &  & \\\hline
3 &  & $2.1\times10^{-7}$ & $7.1\times10^{-13}$ & $7.6
\times10^{-24}$ &
$9.0\times10^{-46}$ & $1.3\times10^{-89}$ &  &  & \\\hline
4 & $4.9\times10^{-12}$ & $1.1\times10^{-21}$ & $5.4
\times10^{-41}$ &
$1.3\times10^{-79}$ & $7.3\times10^{-157}$ &  &  &  & \\\hline
\end{tabular}
\end{table*}

According to (\ref{9}), the pure state $|n,l,R\rangle$ of the
$(n,l)$-universe be an eigenvector of the operator of the
$n$th-level Sfh-spheres number,
$a^{(l)+}_na^{(l)-}_n|n,l,R\rangle=R|n,l,R\rangle$. It is
convenient to represent the mixed states of the $(n,l)$-universe
by the density matrix \be \label{10}
\hat{\rho}^{(l)}_n=\sum^{V^{(l)}_n}_{R=0}w^{(l)}_n(R)
|n,l,R\rangle\langle n,l,R|. \ee The most important observable of
the $(n,l)$-universe is represented by the operator of tetrahedra
(c-points) number for the chc-triangulation of a spatial section
$M^{(l)}_n(R)$,
$\hat{N}^{(l)}_n=V^{(l)}_{n-1}N^{(l)}_n(S)+\left(N'^{(l)}_n(\Sigma)-
N'^{(l)}_n(S)\right)a^{(l)+}_na^{(l)-}_n$, where
$V^{(l)}_{n-1}N^{(l)}(S)$ is tetrahedra number in the ``vacuum''
state $|n,l,0\rangle$ of the $(n,l)$-universe (see Observation 2).
The expectation value of this operator in the mixed state
(\ref{10}) is $\langle N^{(l)}_n\rangle=
\text{tr}\left(\hat{\rho}^{(l)}_n\hat{N}^{(l)}_n\right)$. Its
integer part is equal to the md-volume $V^{(l)}_n$ of the ensemble
${\mathcal M}^{(l)}_n$ (see (\ref{8})). Since this quantity was
obtained as a mean value for all admissible topologies of spatial
sections of the $(n,l)$-universe, it should be treated as the
complexity characteristic of the $T_0$-discrete spacetime (the
definition see in \cite{5}) where only one fundamental interaction
is involved. In the equilibrium (highest probability) state of the
$(n,l)$-universe the observable $N^{(l)}_n$ takes the value
$\simeq V^{(l)}_n$ with the standard relative deviation (mean
relative fluctuation) $\left\{\langle\left(N^{(l)}_n-\langle
N^{(l)}_n \rangle\right)^2\rangle\right\}^{1/2}/\langle
N^{(l)}_n\rangle \simeq 1/\sqrt{V^{(l)}_n}$ with respect to the
binomial distribution $w^{(l)}_n(R)$. This quantity (multiplied by
$2\pi$) we interpret as the dimensionless coupling constant
$\alpha^{(l)}_n:=2\pi/\sqrt{V^{(l)}_n}$ of the unique
$(n,l)$-interaction involved in the most probable state of the
$(n,l)$-universe whose spacetime is built of the ensemble
${\mathcal M}^{(l)}_n$ of the spatial sections topologies. Thus in
our model the values of coupling constants are determined by the
topology fluctuations in the framework of certain finite
ensembles. This hypothesis is supported by the fact that the
cosmological constant in the De Sitter solution (which is used in
inflationary models) is inverse proportional to the square root of
the four-dimensional universe volume \cite{14}. Note that
$V^{(l)}_n$ is the measure of complexity related just to the
spacetime structure of the $(n,l)$-universe, not to its single
spatial section. Hence any coupling constant $\alpha^{(l)}_n$
should be an analogue of the cosmological constant in its proper
$(n,l)$-universe. The values of $\alpha^{(l)}_n$ are given in
Table \ref{tab:3}.

\begin{center}
{\bf Universes with several fundamental interactions}
\end{center}

From the Table \ref{tab:3} one can see that at $t\equiv n-l=0$ the
values of $\alpha^{(l)}_n$ well represent the hierarchy of the
DLEC constants ({\em cf.} the experimental section of the Table
\ref{tab:1}). Thus the family of ensembles of topologies
${\mathcal M}(0)=\left\{{\mathcal M}^{(l)}_n|n-l=0\right\}$ should
correspond to the present state of the universe with five
($n\in\overline{0,4}$) low-energy fundamental interactions. To
describe the states of this compound universe, one should use (as
this is usually done treating compound systems) vectors of the
Hilbert space ${\mathcal H}(0):=\bigotimes_{n-l=0} {\mathcal
H}^{(l)}_n$, the tensor product of all Hilbert spaces related to
the $(n,l)$-universes under the restriction $t=0$. For other
values of $t\in\overline{-4,4}$, the Hilbert space ${\mathcal
H}(t) :=\bigotimes_{n-l=t}{\mathcal H}^{(l)}_n$ includes the
universe states both of the ``past'' ($t<0$) and of the ``future''
($t>0$). Note that due to the restriction $n-l\equiv
t=\text{const}$, the creation and annihilation operators do {\em
not} bring the state vectors from one Hilbert space to another, if
the both are distinct factors in the tensor product of ${\mathcal
H}(t)$; however, $\alpha^{(l)\pm}_n$ perform mapping some state
vectors from ${\mathcal H}(t)$ to ${\mathcal H}(t\pm 1)$, see the
Observations 2 and 3.

From the Table \ref{tab:3} it follows that the number of
$(n,l)$-interactions in the families ${\mathcal
M}(t)=\left\{{\mathcal M}^{(l)}_n|n-l\equiv t=
\text{const}\right\}$ is growing from 1 to 5 when $t$ changes from
$-4$ to 0, while the universe passes four inflationary stages. The
further growth of $t$ from 0 to 4 results in decrease of the
number of interactions to only one, and it is possible to speak on
four ``deflationary'' stages. The initial stage of the universe
with the unique $(0,4)$-interaction is not identical to its final
stage with the unique $(4,0)$-interaction (both the topologies of
the admissible spatial sections and the coupling constants values
are different). We conclude that in our model contains the
unification of fundamental interactions (known from the gauge
theories). It manifests itself here as alternation of the families
of interactions; for example, in the transition from the family
${\mathcal M}(-1)$ to ${\mathcal M}(0)$, the number of
interactions grows from 4 to 5, and in the latter family they are
identified by the values of coupling constants ({\em cf.} the
column $t=0$ in Table \ref{tab:3} with the Table \ref{tab:1}). For
any of these low-energy interactions its natural counterparts can
be traced  both when the quantum number $l$ remains the same, and
when $n$ does not change. {\em E.g.}, the electromagnetic,
$(1,1)$-, interaction has four ``closest counterparts'' in the
higher-energy ensembles ${\mathcal M}(\pm1)$: for $l=1$ these are
$(0,1)$- and $(2,1)$-interactions, and for $n=1$, $(1,0)$- and
$(1,2)$-interactions (see values of their constants in the Table
\ref{tab:3}). Thus we come to a somewhat different unification
approach to fundamental interactions than that known in the gauge
models. The further study of $(1,1)$- (electromagnetic)
interaction shows that its constant after the inflation which
describes the transition from the family ${\mathcal M}(-1)$ to
${\mathcal M}(0)$  changes (due to growth of number of the
homological spheres $\Sigma^{(1)}_1$) from $\alpha^{(1)}_{1\text{
max}}\simeq 1/84.981$ to $\alpha^{(1)}_{1\text{ min}}\simeq
1/190.219$, see the relation (\ref{7}). At present the fine
structure constant has in our universe the value $\alpha\simeq
1/137.036$. It is remarkable that $\alpha^{(1)}_{1\text{ min}}
<\alpha<\alpha^{(1)}_{1\text{ max}}$; moreover, a more profound
observation suggests itself. The most probable value of $\alpha$
(at the equilibrium) calculated according to (\ref{8}), is
$\alpha^{(1)}_{1}=2\pi/ \sqrt{V^{(1)}_{1}}\simeq 1/147.323$. Since
the discrete volume of the universe at the equilibrium is greater
than it is at present ($\alpha^{(1)}_{1} <\alpha$), one has to
conclude that at the present stage the universe should be
expanding, just as it is observed. Strictly speaking, this
observation does pertain to the ``electromagnetic'', {\em i.e.}
$(1,1)$-universe, but it is exactly the electromagnetic radiation
which allows us to detect the Hubble expansion effect. Using other
low energy interactions, it would be more difficult to argue in
this respect, since only the fine structure constant is determined
unambiguously and other dimensionless coupling constants involve
in the existing theory certain rather arbitrary parameters.

Our approach also makes it possible to evaluate the size of
universe as well as the time scales of the stages of its
evolution. To this end note that the counterparts of the
cosmological $(4,4)$-interaction are the $(n,4)$-interactions
($n\in\overline{0,4}$, $l=4$) and $(4,l)$-interactions
(($l\in\overline{0,4}$, $n=4$). It is natural to admit that the
linear size of $(n,l)$-universes can be evaluated by the formula
$R^{(l)}_{n}\sim \sqrt[4]{V^{(l)}_{n}}$. This linear size is
dimensionless. To properly normalize it, admit that at the level
of the unification of four interactions (strong, electromagnetic,
weak, and gravitational ones), {\em i.e.} for $t=n-l=-3$ ($l=4$,
$n=1$), the universe should be of Planckian scales. Then
$r^{(4)}_{1}=k\sqrt[4]{V^{(4)}_{1}}=1Pl=1.63\times10^{-33}cm$.
Admitting the coefficient $k$ to be universal, we evaluate the
linear size of the $(n,l)$-universes as
$r^{(l)}_{n}=kR^{(l)}_{n}$. Just this quantity (in {\it cm}) is
given in the first line of the Table \ref{tab:3}. Note that since
our model takes into account the unification of five interactions
(including the cosmological one), it treats also sub-Planckian
scales; moreover, it predicts the radius of universe to be almost
ten orders of magnitude larger than that of the observable part of
the universe, $R\sim 10^{28}cm$.

Finally, let us summarize the principal results which we came at
in this paper.\\
$\bullet$ A model of universe is constructed where coupling
constants values $\alpha^{(l)}_n$ of the fundamental interactions
are determined by topology fluctuations in the framework of finite
topology ensembles ${\mathcal M}^{(l)}_n$.\\
$\bullet$ The hierarchy of the coupling constants being obtained
in the framework of the topology ensembles ${\mathcal M}(0)$ well
reproduces the experimentally known hierarchy of the low-energy
dimensionless
coupling constants.\\
$\bullet$ The model describes four inflationary stages which
result in the changes of families of the fundamental
$(n,l)$-interactions as well as of the linear scales of the
universe (evaluated via the ``cosmological'' $(n,4)$- and
$(4,l)$-interactions). Four  ``deflationary'' stages are also
predicted which result for the universe in coming to the final
state with only one fundamental $(4,0)$-interaction not being
equivalent to
the initial one with the $(0,4)$-interaction.\\
$\bullet$ All the fundamental coupling constants corresponding to
the most probable states, are explicitly calculated from the
hypothesis that $\alpha^{(l)}_n:=2\pi/\sqrt{V^{(l)}_n}$ where
$V^{(l)}_n$ is the mean discrete volume of the $T_0$-discrete
spaces realized as nerves of chc-triangulations of the spaces of
the ensemble ${\mathcal M}^{(l)}_n$. These triangulations are
compatible with the Milnor representation of the Seifert homology
spheres of which are built all the spaces of all
ensembles ${\mathcal M}^{(l)}_n$.\\
$\bullet$ From the known value of the fine structure constant in
comparison with that calculated in our model, we explain the
universe expansion at the modern stage of its evolution.

\begin{center}
{\bf Acknowledgements}
\end{center}

We are grateful to A.M. Hern\'andez Magdaleno, V.N. Shch\"otochkin
and D.N. Persick for fruitful discussions and friendly help.

\end{document}